\documentclass[conference]{IEEEtran}%
\usepackage{amsmath}
\usepackage{amssymb}
\usepackage{amsthm}
\usepackage{mathtools}
\usepackage{theoremref}
\usepackage{tikz}
\usetikzlibrary{calc,matrix}
\usepackage{cases,setspace,adjustbox}
\usepackage{subfig}
\usetikzlibrary{arrows}
\usepackage{float}

\newcommand{\ud}{\,\mathrm{d}}

\theoremstyle{definition}

\newtheorem{proposition}{Proposition}

\theoremstyle{remark}

\begin{document}
\title{Welfare Analysis of Network Neutrality Regulation}
\author{Rakesh Chaturvedi$^{*}$, Sneihil Gopal$^{\dagger}$ and Sanjit Krishnan Kaul$^{\dagger}$\\
$^{*}$Department of Social Sciences \& Humanities, IIIT-Delhi, India\\
$^{\dagger}$Wireless Systems Lab, IIIT-Delhi, India\\
\{rakesh, sneihilg, skkaul\}@iiitd.ac.in}
\maketitle
\begin{abstract}
Consumers of Internet content typically pay an Internet Service Provider (ISP) to connect to the Internet. A content provider (CP) may charge consumers for its content or may earn via advertising revenue. In such settings, a matter of continuing debate, under the umbrella of net neutrality regulations, is whether an ISP serving a consumer may in addition charge the CPs not directly connected to the ISP for delivering their content to consumers connected to the ISP. We attempt an answer by looking at the problem through the lens of a regulator whose mandate is to maximize the cumulative welfare of ISPs, CPs, and consumers.

Specifically, we consider a two-sided market model, in which a local monopoly ISP prices Internet access to consumers and possibly to CPs as well. The CPs then decide whether to enter a competitive but differentiated market and the consumers decide whether to connect to the ISP. Unlike prior works, we model competition between the CPs together with consumer valuation of content and quality-of-service provided by the ISP. We do so by using a novel fusion of classical spatial differentiation models, namely the Hotelling and the Salop models, in addition to simple queue theoretic delay modeling. Via extensive simulations, we show that the equilibrium in the non-neutral setting that allows an ISP to charge a CP welfare-dominates the neutral equilibrium.

\end{abstract}
\IEEEpeerreviewmaketitle
\section{Introduction}
\label{sec:introduction}
Network neutrality prohibits an ISP from discriminating between packets of similar type of content from different providers by either pricing their transit differently or by providing them different qualities of service. As has been well articulated by~\cite{coucheney2013impact}, ISP(s) that are not directly connected to CPs and provide a transit to the content generated by the CPs have argued that CPs are cornering good amounts of dividends on an investment that was made by the ISPs. While ad-revenues are rising for the CPs, the transit prices that are a source of revenue for such ISPs are decreasing. The ISPs would thus like the CPs to pay for transit of their content or suffer loss in quality-of-service.

On the other hand, the proponents of net neutrality point to the detrimental effect that a non-neutral regime can have on the CPs and by implication on the consumers. For example, as pointed out in~\cite{economides2012network}, a non-neutral regime could impact innovation at the edge of the network, as it could now be influenced by the ISPs that constitute the core of the network. Allowing discriminatory pricing or provisioning of quality-of-service may enable an ISP to adversely impact access of content by consumers of some among competing CPs.

In this work, we will analyze net neutrality from the point of view of a regulator whose mandate is to maximize the sum welfare of all consumers, CPs, and the ISPs. Note that the larger the number of CPs providing similar content, the higher is the intensity of competition amongst providers. More competition must increase consumer welfare. However, too much competition amongst providers may squeeze their revenues so much that they end up making negative returns on the cost of their investment to set up business. Our regulator can't merely maximize consumer welfare but must contend with the impact of too much competition on the welfare of CPs. A natural line of inquiry is then to include such aspects in any welfare analysis of net neutrality regulation. For example, models used for analysis must capture competition in the provider market. However, we are not aware of any previous work that looks into this direction. 

By nature, the ISP operates in a two-sided market. It sells network access to both consumers and CPs. Consumers can then choose to consume content from providers. Our model is therefore a two-sided market model in the style of~\cite{economides2012network}. In this paper, we restrict ourselves to analyzing the case of a monopolist ISP.  We model the interaction of the ISP, the CPs and the consumers as a multi-stage game. The ISP moves first to set access prices for both consumers and CPs. The CPs move next making a binary choice of whether to enter the market. Finally, consumers choose whether to connect to the ISP. This structured interaction via a multi stage game permits an analysis of the influence of the pricing decisions of the ISP on the size of the CP market and the size of the consumer market. Another feature of our modeling is that CPs provide all their content free to consumers\footnote{This is motivated by the providers' business model and agrees with the literature.}; so their only source of revenue is advertising. 

In our model, all CPs provide similar content. However, differences in consumer tastes imply that even at the same price a consumer may prefer a certain CP over another. That is the content providers are not perfect substitutes and their content offerings are horizontally differentiated~\cite{tirole1988theory}. This guides how the providers arrange themselves among the consumers of the ISP. Similarly, consumers have non-price related preferences with regards to the ISP. Additionally, we make the reasonable assumption that it is more difficult for a consumer to find an alternate ISP than it is to find an alternate CP. 

To capture the aforementioned we use two workhorse models -- namely the linear city model due to \emph{Hotelling}~\cite{hotelling1929} and the circular city model due to \emph{Salop}~\cite{salop1979monopolistic} -- from the industrial organization literature~\cite{tirole1988theory}. Note that a consumer consumes the two products of network access sold by the ISP and content sold by a CP. Technological constraints imply that a consumer must first have network access to subsequently consume content. We, therefore, face two product dimensions -- one for the ISPs and one for the CPs. To capture these, we use a novel way to model our market, which we will call the \emph{Hotelling-Salop} model, where in we employ the \emph{Hotelling} line for the ISP dimension that determines how many consumers connect to the ISP and then let the resulting segment of that line to be the circumference of the \emph{Salop} circle on which CPs align to provide content to the consumers.

In our model, the ISP has a fixed amount of bandwidth that it must use to service requests of content made by its consumers. Thus one would expect the quality-of-service (average delay in this paper) obtainable by any given consumer to be impacted by the demands for content made by the other consumers. To capture this fact, as in~\cite{pil2010net}, we use the M/M/1 queueing model, which assumes that the aggregate consumer demand at the ISP is modeled as a Poisson process and that the time it takes for the ISP to service a consumer's request for content is well modeled by an exponential distribution. The resulting average delay is treated as a cost borne by every consumer. While the M/M/1 model is simplistic, it allows us to capture the important fact that average delay (cost) suffered by a consumer increases -- slowly, when the aggregate demand is small and doesn't congest the available bandwidth, and rapidly as the aggregate demand makes the network congested -- as the number of consumers of the ISP increase.

Finally, in our model, as in~\cite{economides2012network}, network neutrality is defined as the constraint that the ISP cannot charge the CPs. 

\thref{WelfareCompare} in Section~\ref{sec:Welfare} summarizes our main result, which is that the non-neutral equilibrium dominates the net neutral equilibrium. As is borne out by our analysis, this is because welfare depends on the two variables of consumer and provider market size and not on the prices charged of the consumers and providers. The prices do not matter for welfare as they end up being mere transfers from one player to another. A switch from a non-neutral to a neutral regime leads to welfare losses on both counts -- a decrease in the consumer market size that it entails decreases the economic value generated by consumers (net of any gains achieved from reduced congestion costs); and an explosive increase in the provider market size (way above the welfare optimum) imposes costs that far outweigh any benefits to consumers from increased diversity of content providers.


The rest of this paper is organized as follows. Section \ref{sec:related} provides a limited overview of the growing literature on net neutrality. In Section~\ref{sec:spatialModels}, we provide a brief on the use of spatial models to model competition and consumer preferences. In Section \ref{sec:model}, we set up the model and pursue the equilibrium analysis under non-neutral and neutral regimes. We then go on to show how to do a welfare analysis of equilibria and also how to compute a welfare optimum. Computational analysis is carried out in Section \ref{sec:analysis}. Lastly, we make a few concluding remarks in Section \ref{sec:conclusion}.
\section{Related Work}
\label{sec:related}
The contours of the debate on net neutrality are by now well known and well articulated in works like~\cite{kramer2013net} and~\cite{greenstein2016net}. As we state in the introduction, our modeling objective is to compare in welfare terms the equilibriums in neutral and non-neutral regimes. Non-neutrality introduces many feasible strategies for the ISP, including two-sided pricing and the possibility of prioritization of network lanes. The literature has progressed on both these dimensions. In the direction of work on lane prioritization, in~\cite{ma2017paid} it is found that paid prioritization can be welfare-superior to net neutrality. In~\cite{zou2017optimal} the authors investigate optimal service differentiation for an ISP subject to capacity constraints in an optimal control framework. In~\cite{gharakheili2016economic} non-neutrality is studied in an economic model that allows lane pricing by ISP and revenue generation for CP as well as QoE improvement for the consumers. The economic benefits for ISP and CP are analyzed using traffic traces from real networks. However, lane prioritization is not the focus of our work. In doing this, our work is closer in spirit to~\cite{economides2012network} where too authors abstract from lane prioritization issues.

In the context of a two-sided pricing model, wherein ISP charges both consumers and content providers,~\cite{altman2011network} studies non-neutrality in an economic model including an ISP, a CP, consumers and advertisers and shows that side-payments are beneficial for both the ISP and the CP. In~\cite{andrews2013economic} the interaction between an ISP and a CP is modeled as a Stackelberg game with the objective of maximizing the total profit of both players. In~\cite{coucheney2013impact} the authors study the impact of competition between ISPs on network neutrality. Authors in~\cite{kamiyama2014feasibility} analyze a model with competing ISPs, multiple non-competing CPs and a fixed number of consumers and study the feasibility of imposing side-payments on CPs and its effect on the revenue of both ISP and CP. Other works such as~\cite{ma2013public} and~\cite{mitra2017} study the impact of net-neutrality on consumer utility and investments in the network focusing on incremental bandwidth allocation and caching, respectively. 

While prior works consider competing ISPs and multiple non-competing content providers, to the best of our knowledge, our work is the first attempt to study non-neutrality with a focus on competition issues in the provider market for a particular content type. We are interested in how the potential of two-sided pricing introduced by non-neutrality affects the competitive landscape of the provider market, as well as the consumer market base, with consequences for welfare. 

In terms of modeling choices, our work is closely related to~\cite{economides2012network} and~\cite{pil2010net}. However, while we study competition in the content provider market, authors in~\cite{economides2012network} model content providers as many independent monopolies. Also, in our work consumers face a congestion cost that is absent in~\cite{economides2012network}.

\section{Modeling Consumer Choice and Competition}
\label{sec:spatialModels}
We need to model consumer preference for competing content providers and for the ISP connection. We will think of both content and the ISP connection as economic goods of which the consumers are the buyers and the content providers and the ISP are respectively the sellers.

Consumer preference is often guided by a mix of price and a myriad of non-price factors. To exemplify, while both Flipkart and Amazon are content providers in the e-commerce space in India and they have similar transaction costs, some Indian consumers may prefer one over the other. Such non-price factors are modeled in the industrial organization~\cite{tirole1988theory} literature using \emph{spatial models of product differentiation}. Spatial models use an abstraction of a single dimension -- a \emph{Hotelling} line or the perimeter of a \emph{Salop} circle -- to capture all non-price factors. Buyers and sellers are identified by locations on this dimension. A buyer's distance from a seller on the single dimension captures the non-price aspects of the buyer's preference for the seller. On this dimension, for our example, a consumer that prefers Flipkart will be closer to Flipkart than to Amazon.

In addition to the above, spatial models have a product differentiation parameter that captures the ease with which a buyer can substitute between offerings of competing sellers. The smaller this parameter, the easier it is for the buyer to substitute one seller for another. Specifically, the cost incurred by a buyer to change its position (relative to the sellers) on the single dimension is smaller.

To summarize, we use a richer model of price competition with differentiated products that is closer to reality than the simplistic model of price competition with no differentiation. We remark that in~\cite{coucheney2013impact} an alternative price based discrete choice model is used to address many of the same modeling issues. However, we find the spatial models attractive for their relative simplicity.
\section{Basic Model}
\label{sec:model}
There is a monopoly ISP and many potential CPs for whom this ISP is not the originating ISP. Introduce the interval $[0,1]$ as the \emph{Hotelling} line of the standard horizontal product differentiation model for the ISPs. Let the ISP be located at the left extremity, location $0$, of this line. There is a continuum of consumers uniformly distributed on the Hotelling line. Let the product differentiation parameter for the ISP be normalized to $1$. Non-price factors therefore impose a cost of $x$ on a consumer at location $x$ on the Hotelling line.

Let $t \in (0,1)$ be the product differentiation parameter for the CPs. Note that we restrict $t<1$. This is to capture the fact that it is easier to substitute between CPs than it is to substitute between ISPs. Let $y$ be the distance travelled by a consumer to connect to a CP. This imposes a cost of $ty$ on the consumer. Consumers cannot connect to the CPs without first connecting to the ISP. The CPs space themselves equidistantly on a Salop circle whose perimeter, as we detail soon, is determined by the consumers that choose to connect to the ISP. 

All CPs provide free content to the consumers; their only source of revenues is from advertising. Let $r$ be the advertising revenue per consumer, same for all CPs. Let $d$ be the access price charged by the ISP to the consumer. Let $f$ be the fixed cost of entering the provider market for any CP.

Consumers pay a congestion cost, which is the average time the ISP takes to fulfill a consumer's request for content. To capture this, we model the ISP as a M/M/1~\cite{gross1998fundamentals} service facility. Specifically, following~\cite{pil2010net}, we assume that the aggregate demand for content from consumers created at the ISP is well modeled by a Poisson process of rate $\lambda$ per unit length of the single dimension containing consumers connected to the ISP. Further, the time taken by the ISP to service a consumer's request for content is exponentially distributed with a mean service time of $1/\mu$. We assume $\lambda < \mu$, which is required for the average time to fulfill a request to stay bounded. Finally, consumer requests are serviced by the ISP in a first-come-first-served manner. This means that a request that arrives at the ISP must wait for requests that arrived earlier to finish obtaining service from the ISP. 

Consumer monetary valuation of the total requested content is $v$. In the non-neutral regime, the ISP can charge the CPs for servicing requests for their content at an access price of $a$. In contrast, the neutrality regime prohibits the ISP from charging the CPs a price when it is not the originating ISP for them, which we assume is the case here. The two regimes therefore induce different market games through which the entities interact.


\subsection{Non-neutral Regime}
\begin{figure}
	\centering
	\includegraphics[width=0.9\columnwidth]{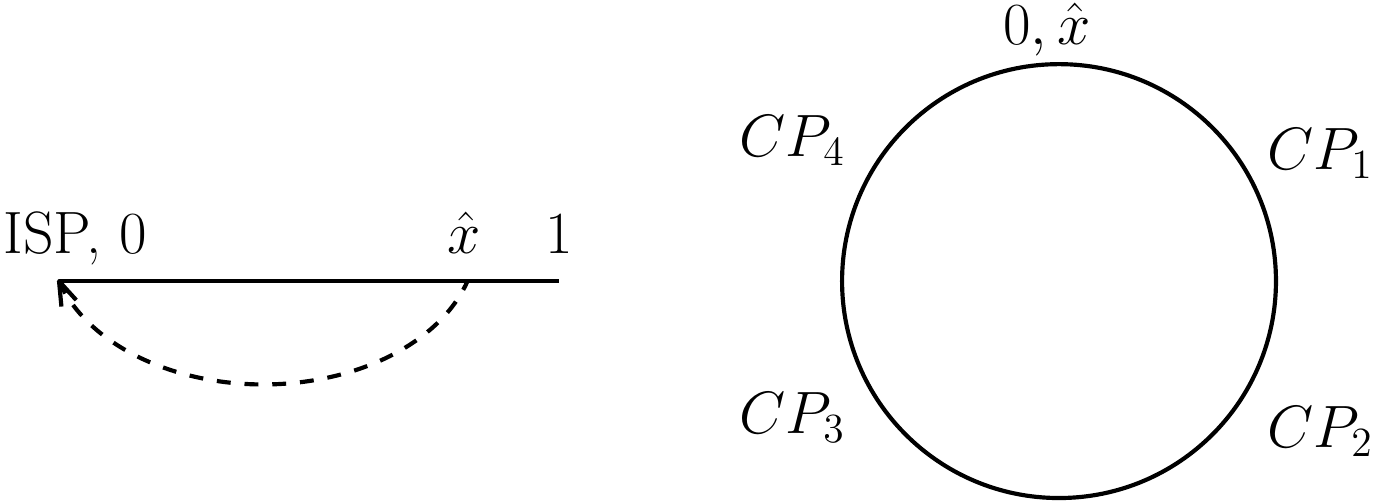}
	\caption{\small{Left Panel shows the \emph{Hotelling} line for the ISP dimension. Right panel shows the folded \emph{Hotelling} line segment of length $\hat{x}$ as the \emph{Salop} circle with the same circumference. It also shows the symmetric alignment of CPs when $n=4$ for when the consumer at $\hat{x}$ travels a distance $\hat{x}/(2n) = \hat{x}/8$ to reach a CP.}}
	\label{fig:ModFig}
\end{figure}
In the non-neutral regime, the ISP, the CPs and the consumers interact through the following \emph{multi-stage game}:  
\begin{enumerate}
\item ISP chooses the access prices: $a$ for CPs and $d$ for consumers.
\item Consumers then choose whether to connect to the ISP.
\item CPs who are considering entering the market then simultaneously decide whether to enter or not. Suppose $n$ is the number of CPs who chose to enter. The CPs locate equidistantly\footnote{We follow \cite{salop1979monopolistic} in making the assumption of symmetric alignment of CPs on the Salop circle to focus on the entry decisions of CPs.} on a \emph{Salop} circle whose perimeter is the length $\hat{x}$ of the initial segment of the \emph{Hotelling} line on which consumers connect to the ISP.
\item Consumers who have connected to the ISP then connect to a CP and consume content. 
\end{enumerate}

We specify utility of a consumer who travels a distance $x$ on the \emph{Hotelling} line to connect to the ISP and distance $y$ on the \emph{Salop} circle to connect to the CP as
\begin{equation}\label{eqn:consumerUtil}
u = v - x  - ty - d - \frac{1}{\mu- \lambda \hat{x}},               
\end{equation}
where $\lambda \hat{x}$ is the aggregate rate of request for content from the length $\hat x$ of consumers. The last term in the expression for consumer utility $u$ is the congestion cost faced by a consumer. It is the resulting expected time the ISP takes to fulfill a consumer's request (called waiting time in~\cite{gross1998fundamentals}), given the demand $\lambda \hat{x}$.


We specify ISP's profit as $\pi(d, a) = d \hat{x} + na$. We specify profits of $\text{CP}_i$ as $\pi_i = r D_i - a - f$, where $D_i$ is the demand that $\text{CP}_i$ faces. 

To determine this demand $D_i$ we must decide how the CPs are aligned on the Salop circle relative to the marginal consumer at $\hat{x}$ who connects to the ISP. This is important for it determines the distance traveled by this consumer on the Salop circle and therefore affects the location $\hat{x}$ on the Hotelling line. Two extreme cases are possible. In one, illustrated in Figure~\ref{fig:ModFig}, the marginal consumer travels a distance $\frac{\hat{x}}{2n}$; in the other, the consumer travels no distance, that is a CP locates itself at $\hat{x}$. %
One could introduce an extra parameter $\beta \in [0,1]$, and specify the distance traveled by the marginal consumer as $\frac{\beta \hat{x}}{2n}$. The extreme cases correspond to the values $\beta = 1$ and $\beta = 0$. We will, however, choose $\beta = 1$ in our analysis for two reasons. First, any $\beta \in (0, 1]$ retains the structural feature that demand on either side (whether $\hat{x}$ or $n$) responds to prices ($d$ and $a$) on both sides. Therefore $\beta = 1$ carries the flavor of analysis of any $\beta \in (0,1)$. Second, as will be clearer from analysis that follows, $\beta = 0$, kills any dependence of $\hat{x}$ on $a$ and in that sense is a special case.

 
To summarize, the parameters of the model are the consumer valuation $v$, the advertising revenue $r$, the normalized product differentiation parameter $t$, fixed CP costs $f$, the rate of content request $\lambda$, and the service rate $\mu$. The choice variables are the prices $a$, $d$, the entry decisions of the CPs, and the consumer decisions with regards to connecting to the ISP and the CPs. The solution concept used to analyze the model is subgame perfect Nash equilibrium. We will use the backward induction algorithm to find subgame perfect Nash equilibria of the multi-stage game.


Suppose we are in the subgame in which the mass of consumers who chose to connect to the ISP is $\hat{x}$, the number of CPs who chose to enter is $n$ and they place themselves symmetrically on the \emph{Salop} circle whose perimeter is $\hat{x}$. So $CP_i$ faces a demand $D_i = \frac{\hat{x}}{n}$ with consequent profits
\begin{equation}
\pi_i = r \frac{\hat{x}}{n} - f - a.
\end{equation}
Under free entry, the optimal number of CPs who enter is determined by the zero-profit condition yielding
\begin{equation}\label{CPDemand}
n = \frac{r \hat{x}}{a+f}.
\end{equation}
Ceteris paribus, an increase in ISP's access price $a$ to CPs just adds to their fixed costs and reduces the number of CPs that enter the market. Similarly, an increase in total demand $\hat{x}$ encourages entry.

The marginal consumer at $\hat{x}$ travels a distance $\frac{\hat{x}}{2n}$ and is indifferent between connecting to the ISP and not connecting to it, that is it has a utility of $0$. Setting the utility $u$ in~(\ref{eqn:consumerUtil}) to zero, we can write 
\begin{equation*}
v - \hat{x} - \frac{t \hat{x}}{2n} - d - \frac{1}{\mu- \lambda \hat{x}} = 0.
\end{equation*}

Using $n$ from Equation~(\ref{CPDemand}) in the above equation, we have 
\begin{equation}   \label{ConDemand}
v - \hat{x} - \frac{t (a+f)}{2r} - d - \frac{1}{\mu- \lambda \hat{x}} = 0.
\end{equation}
Let $\hat{x}(d, a)$ solve Equation (\ref{ConDemand}). Then 
\begin{align}
\hat{x}(d, a)  &= \frac{1}{2\lambda}\left(\mu + \lambda \big(v-d - \frac{t (a+f)}{2r} \big)\right.\nonumber\\
&\left.- \sqrt{[\mu - \lambda \big(v-d - \frac{t (a+f)}{2r} \big)]^2 + 4 \lambda}\right). \label{Demand1}
\end{align}
Further, using~(\ref{CPDemand}), we can calculate the number of CPs
\begin{align}
n(d, a) &= \frac{r \hat{x}(d, a)}{a+f}.  \label{Demand2}
\end{align}
So the total demand on either side is decreasing in both the access prices. 

The ISP chooses the access prices $d$ and $a$ to maximize its profits subject to the constraints that the resulting consumer demand is nonnegative and the demand on the CP side is at least 1. Formally it faces
\begin{align*}
[OPT1] \quad \quad  \max_{d, a} \quad \pi(d, a) &= d \hat{x}(d, a) + a n(d,a)\\
\text{subject to } \quad 0 \le \hat{x}(d, a) &\le 1,\\
n(d,a) &\ge 1.
\end{align*}
Define $e(\hat{x}(d, a), d) := - \frac{\ud \hat{x}(d, a)/ \hat{x}(d, a)}{\ud d/ d}$ to be the elasticity of consumer demand $\hat{x}(d, a)$ with respect to $d$. Similarly, define $e(\hat{x}(d, a), a)$ as the elasticity with respect to $a$. Let $e(n(d, a), d)$ and $e(n(d, a), a)$ be the corresponding elasticities of the number of CPs. Then an unconstrained optimum is a solution to the following pair of equations in $(d, a)$.
\begin{align}
e(\hat{x}(d, a), d) + \frac{an}{d \hat{x}} e(n(d, a), d) &= 1,\label{FOC_d}\\
e(n(d, a), a) + \frac{d \hat{x}}{an} e(\hat{x}(d, a), a) &= 1.\label{FOC_a}
\end{align}
Equations (\ref{FOC_d}) and (\ref{FOC_a}) can be interpreted as follows. If the monopolist ISP wants to increase $d$, he faces the traditional margin vs volume tradeoff\footnote{A profit maximizing monopolist facing a downward sloping demand curve faces a tradeoff when choosing an optimal price. While a higher price implies he makes more on the units that he sells (this is the margin effect), it also implies a lower sales volume (this is the volume effect) because a higher price leads to a lower demand. }. That is, he loses some consumers but makes more on the existing consumers. However, since the ISP is operating in a two-sided market, an additional effect arises via network externalities. A decrease in consumer market size makes entry unattractive for CPs. So the provider market size decreases which reduces the ISP's revenues from that market\footnote{That the aforementioned is indeed an interpretation of Equation (\ref{FOC_d}) is established by looking at the more fundamental first order condition with respect to $d$, which is  
\begin{align}
&\frac{\ud \pi}{\ud d}= \underbrace{\hat{x}}_{\text{margin effect}} + \underbrace{d \frac{\ud \hat{x}}{\ud d}}_{\text{volume effect}} + \underbrace{a \frac{\ud n}{\ud d}}_{\text{network externalities effect}} = 0.\nonumber
\end{align}
Equation (\ref{FOC_d}) is equivalent to the above equation under the definition of elasticities.}. Similarly, if the ISP wants to increase $a$, in addition to facing the traditional margin vs volume tradeoff in the provider market, it has to consider that increasing $a$ also decreases the consumer market size which has a second order effect on further reducing volumes in the provider market. The optimal access prices $(d, a)$ balance these effects simultaneously. 

\subsection{Net Neutrality Regime}
Net neutrality regulation constrains the access price $a$ that the ISP can charge CPs to be zero. In the neutrality regime, the ISP, the CPs and the consumers interact through the same multi-stage game as in the non-neutral regime but with the constraint $a = 0$.    

The ISP faces the following optimization problem for choosing the access price to consumers
\begin{align*}
[OPT2] \quad \quad \max_{d} \quad \pi(d, a=0) &= d \hat{x}(d, a=0)   \\
\text{subject to } \quad 0 \le \hat{x}(d, a=0) &\le 1,   && (C_1)\\
n(d, a=0) &\ge 1,     && (C_2)
\end{align*}
where $\hat{x}(d, a=0)$ and $n(d, a=0)$ are found by setting $a = 0$ in Equations (\ref{Demand1}) and (\ref{Demand2}).

The unconstrained optimum of $[OPT2]$ is given by the value of $d$ at which the standard margin vs volume tradeoff in the consumer market is resolved optimally at
\begin{equation}   \label{NeutralOpt}
e(\hat{x}(d, 0), d) = 1
\end{equation}
and is the solution if it lies in the constraint set. The corresponding demands $\hat{x}(d, a=0)$ and $n(d, a=0)$ on both sides can be computed by setting $a=0$ and $d$ at its optimal value in Equations (\ref{Demand1}) and (\ref{Demand2}).

\subsection{Welfare}
\label{sec:Welfare}
The model we are studying is a transferable utility model. This implies that the cardinal welfare $W$ at any outcome is defined to be the sum total of consumers' surplus ($CS$), profits ($\pi_i$) of all CPs who enter and profit $\pi$ of the ISP. We have these expressions
\begin{align}
\pi &= d \hat{x} + na,    \nonumber  \\
\sum_{i=1}^n \pi_i &= r \hat{x} - nf - na,  \nonumber \\
CS &= \int_{0}^{\hat{x}} \big(v-d - \frac{1}{\mu- \lambda \hat{x}} -x \big) \ud x - \nonumber\\
&\quad n \int_{0}^{\hat{x}/n} ty \ud y  = \big(v-d-\frac{1}{\mu- \lambda \hat{x}} \big) \hat{x} - \frac{\hat{x}^2}{2} - \frac{t \hat{x}^2}{n}, \nonumber \\
W(\hat{x}, n) &= \pi + \sum_{i=1}^n \pi_i + CS = (v+r) \hat{x} -  \frac{\hat{x}^2}{2} - \frac{t \hat{x}^2}{n}\nonumber\\ &\quad - nf - \frac{\hat{x}}{\mu- \lambda \hat{x}}.  \label{Welfare}
\end{align}
The access prices cancel out because these are mere transfers from one party to another. The overall welfare in an equilibrium depends on the consumer market size and the number of CPs that populate the provider market. If the consumer market size $\hat{x}$ increases, consumer surplus and advertising surplus increase on account of more consumers being served but welfare decreases on account of two factors -- first, an increase in consumer market size increases the heterogeneity in consumer preferences which causes the product diversity along the ISP and the CP dimension to decrease relatively; second, an increase in consumer market size also increases the congestion cost for everyone. Similarly, an increase in the number of CPs who enter increases consumer surplus because there is more product diversity on the CP dimension for consumers but imposes greater fixed costs on the society.

As is standard in economic analysis, we compute the welfare optimum as a solution to the following problem.
\begin{align*}
[OPT3] \quad \quad \max_{\hat{x}, n} \quad & W(\hat{x}, n)   \\
\text{subject to } \quad 0 \le \hat{x} &\le 1,   \\
n &\ge 1.     
\end{align*}
The solution to $[OPT3]$ gives the size of consumer and provider market that is best from society's point of view when we have taken everyone's interest into account. 
\section{Computational Analysis}   
\label{sec:analysis}
A look at equations (\ref{FOC_d}), (\ref{FOC_a}) and (\ref{NeutralOpt}) and the corresponding expressions for elasticities in the Appendix suggests that a meaningful theoretical analysis does not look feasible. We, therefore, do a computational analysis of $[OPT1]$, $[OPT2]$ and $[OPT3]$. Recall that our objective is to determine whether equilibrium welfare in one regime is higher than in the other.

\begin{figure*}[t!]
\begin{center}
\subfloat[]{\includegraphics[width=0.33\linewidth]{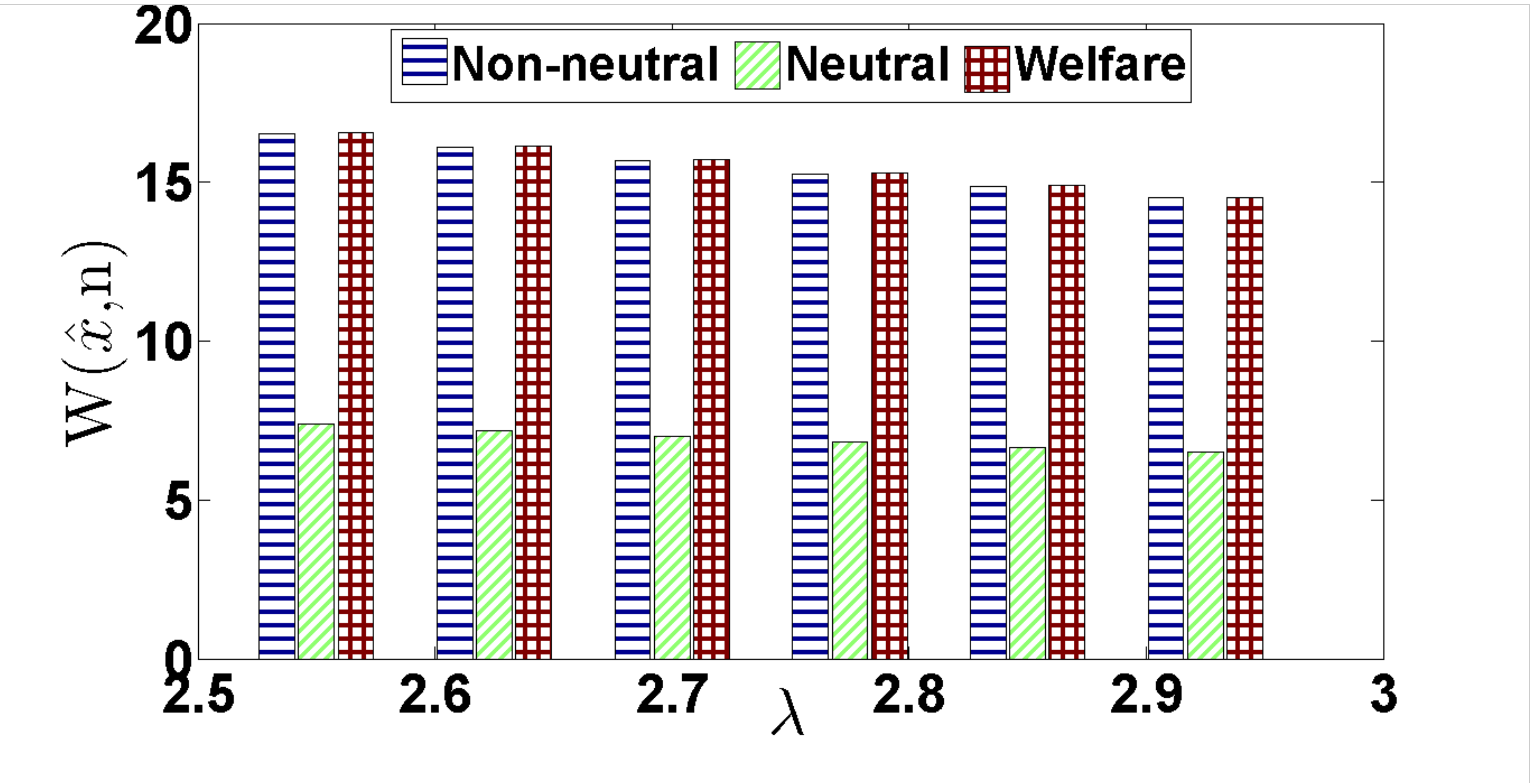}\label{fig:Wel_v_eq_r}}
\subfloat[]{\includegraphics[width=0.33\linewidth]{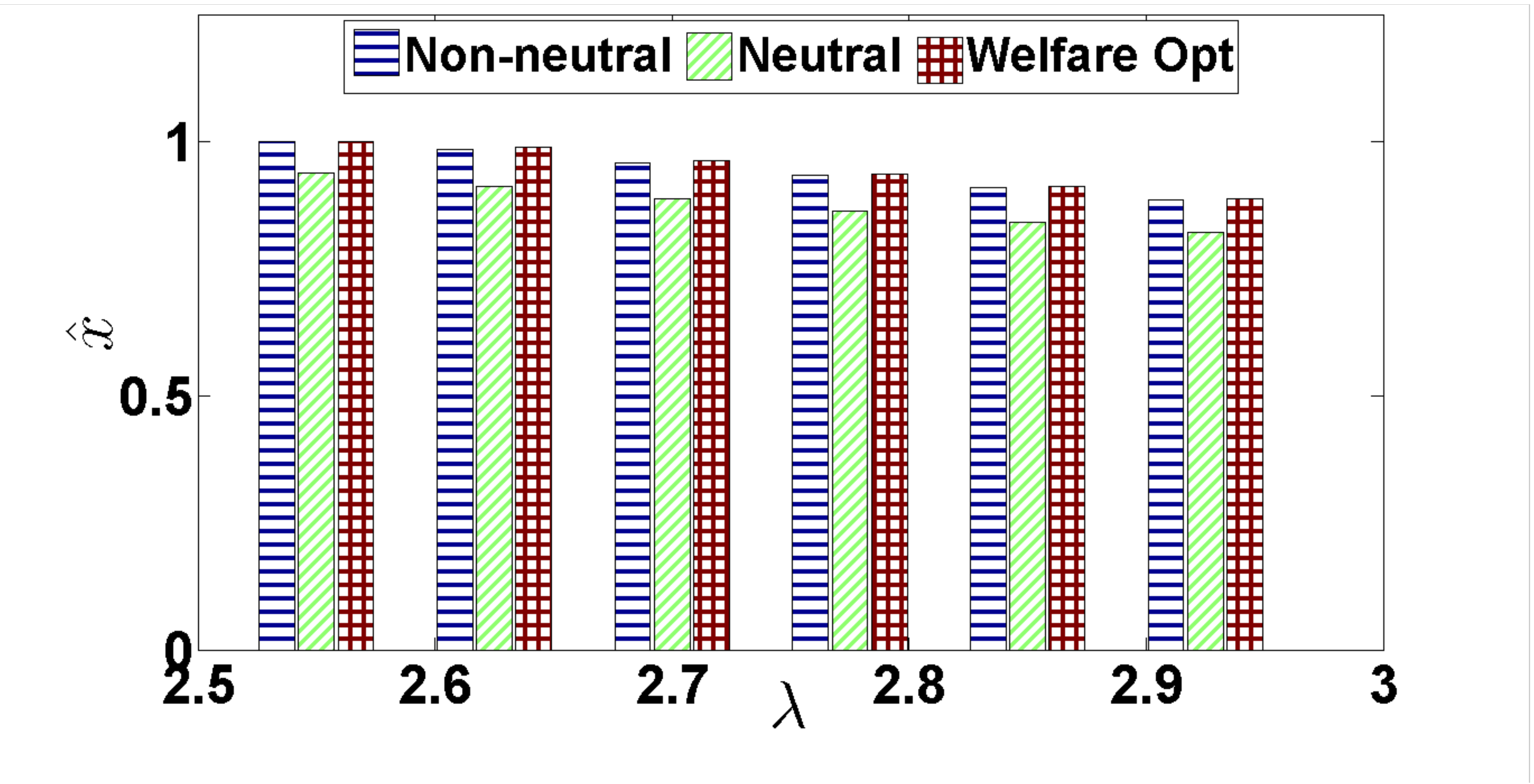}\label{fig:Con_v_eq_r}}
\subfloat[]{\includegraphics[width=0.325\linewidth]{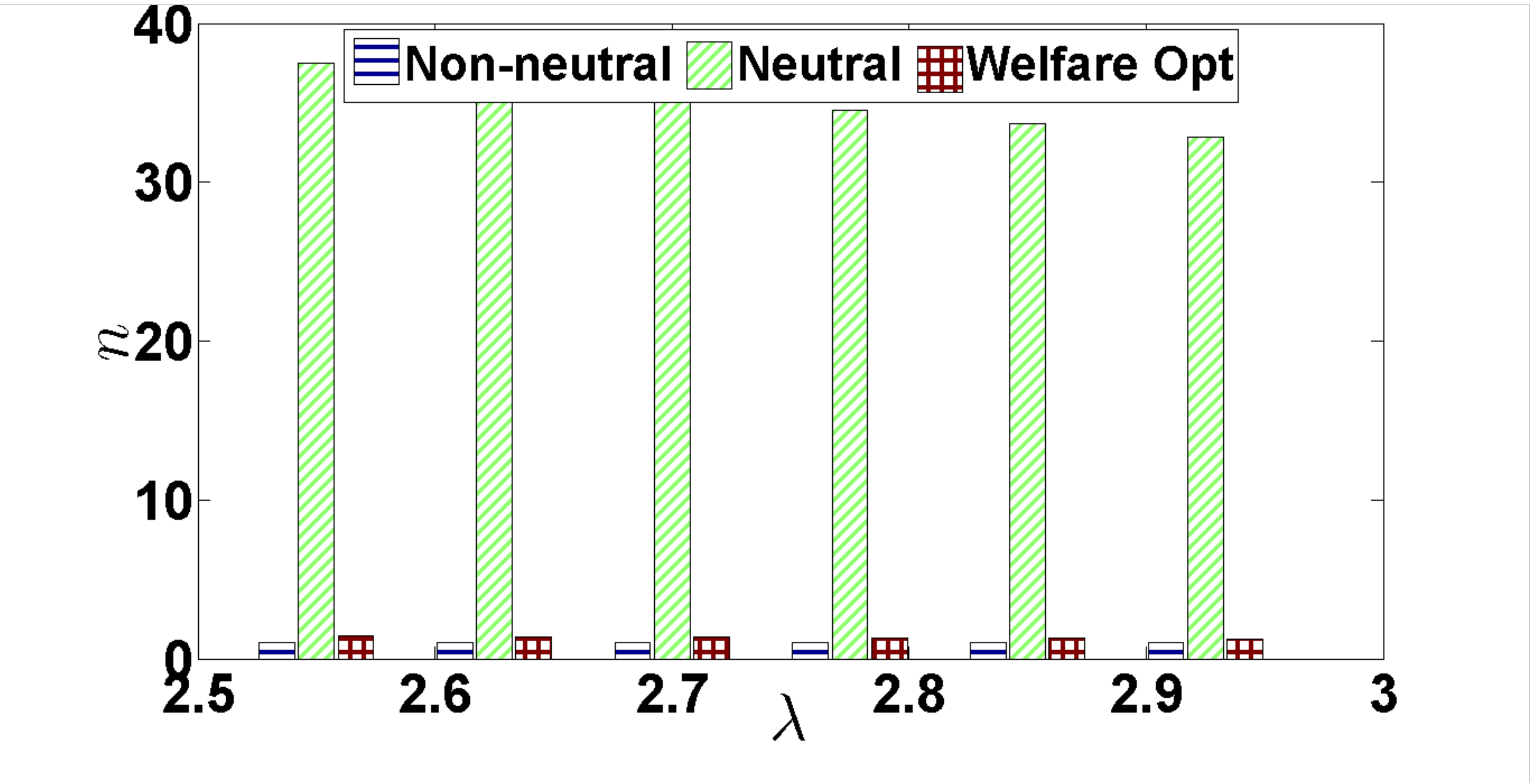}\label{fig:CP_v_eq_r}}\\
\subfloat[]{\includegraphics[width=0.33\linewidth]{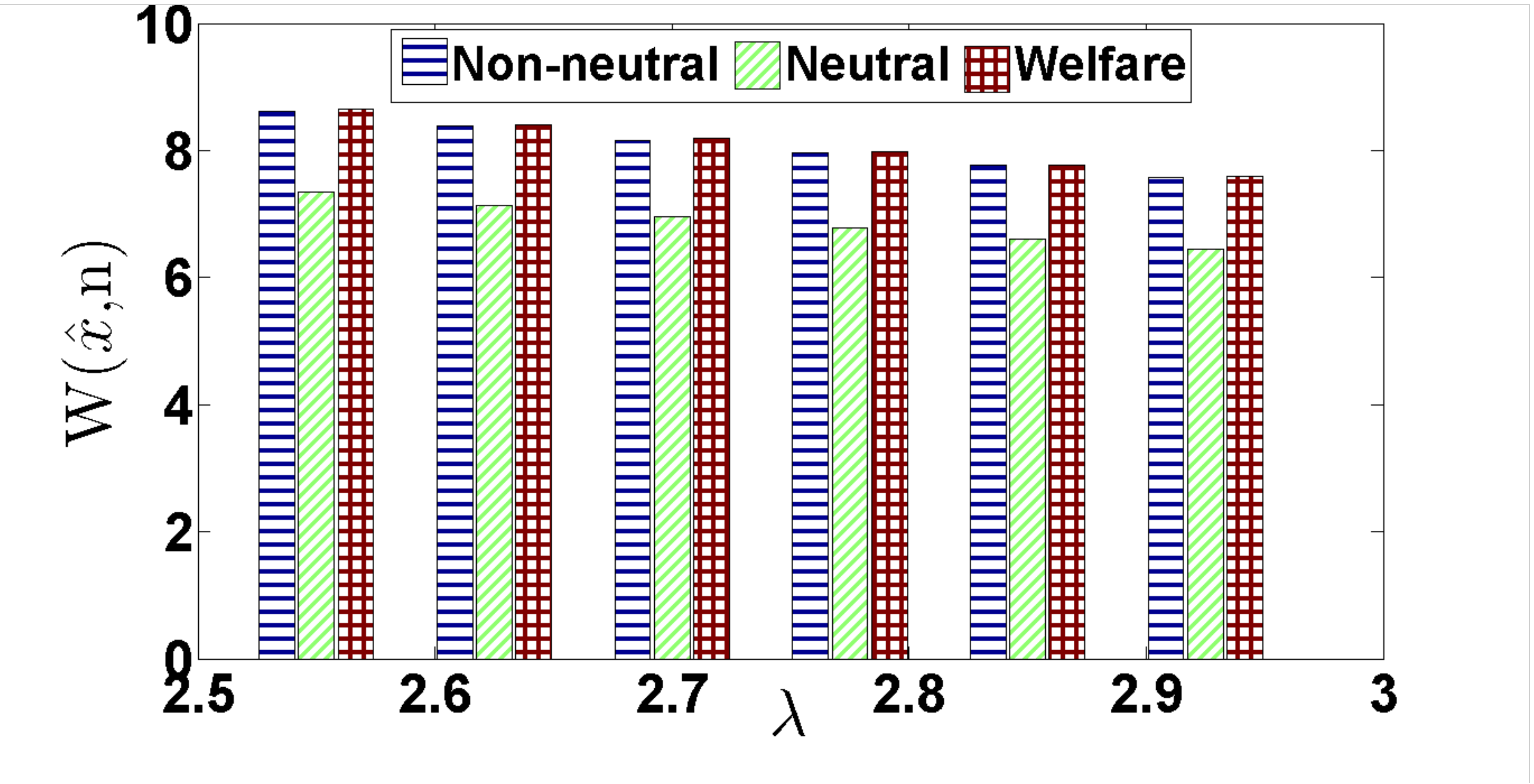}\label{fig:Wel_v_gr_r}}
\subfloat[]{\includegraphics[width=0.33\linewidth]{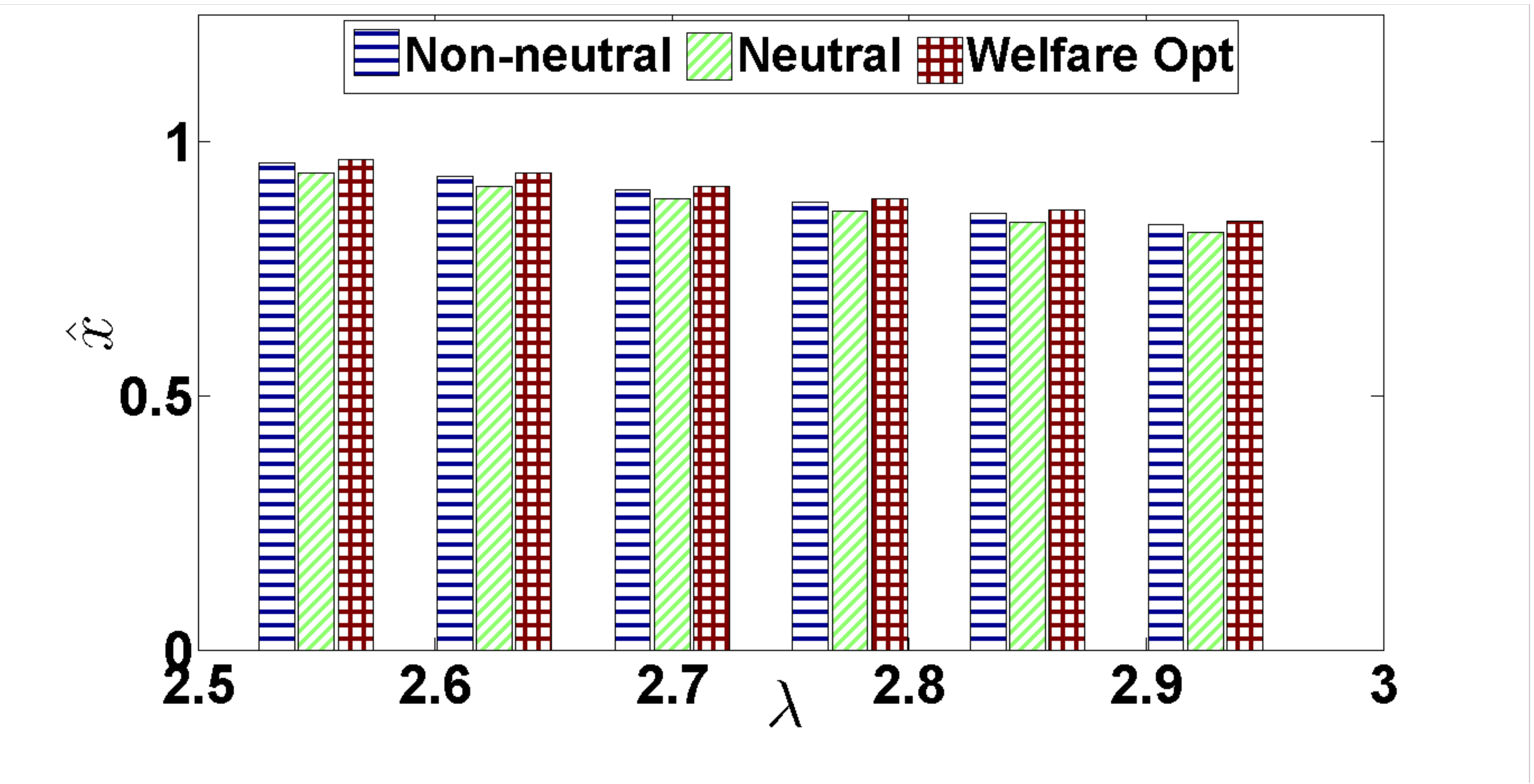}\label{fig:Con_v_gr_r}}
\subfloat[]{\includegraphics[width=0.325\linewidth]{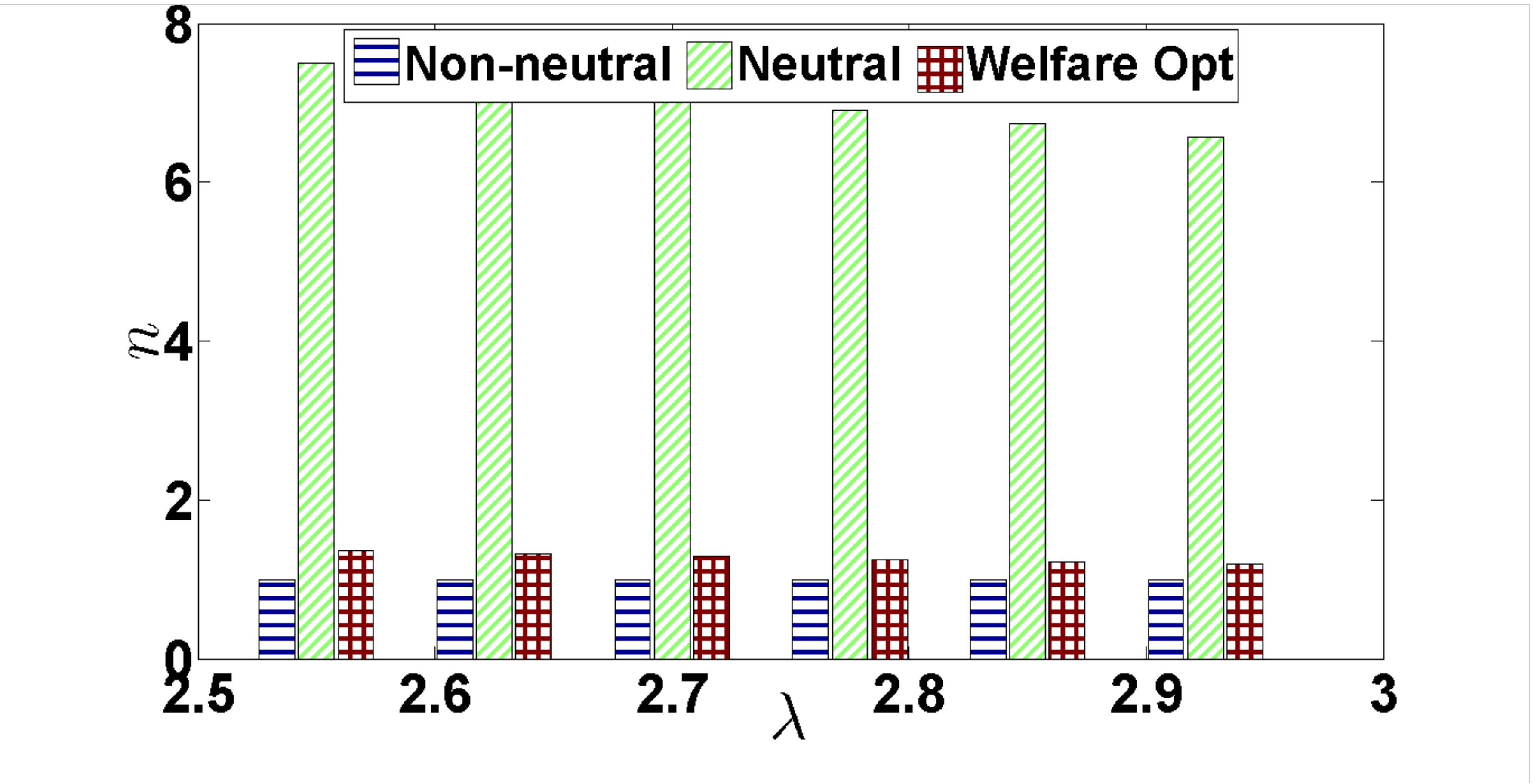}\label{fig:CP_v_gr_r}}\\
\caption{\small{(a) Equilibrium Welfare (b) Consumer Market Size and (c) Provider Market Size for $v=r=10$, and (d) Equilibrium Welfare (e) Consumer Market Size and (f) Provider Market Size for $v=10$ and $r=2$, for the cases of Non-neutral Equilibrium, Neutral Equilibrium and Welfare Optimum. Other parameters used for these results are $t = 0.5, f = 0.25,\text{ and }\mu = 3$.}}
\label{fig:VandR}
\end{center}
\end{figure*}


Extensive computational analysis reveals that we get an unambiguous welfare ranking of the non-neutral and neutral equilibrium for a large range of parameter values. Specifically, \emph{the non-neutral equilibrium welfare-dominates the neutral equilibrium and is close to the welfare optimum}. Figure~\ref{fig:VandR} demonstrates this using example results for when the consumer valuation $v$ is equal to the advertising surplus $r$ (Figures~\ref{fig:Wel_v_eq_r}-\ref{fig:CP_v_eq_r}) and when $v > r$ (Figures~\ref{fig:Wel_v_gr_r}-\ref{fig:CP_v_gr_r}). Qualitatively, the results (not shown) are similar for when $v < r$.


Figures~\ref{fig:Con_v_eq_r} and~\ref{fig:Con_v_gr_r} compare the consumer market size $\hat x$ under the neutral equilibrium, non-neutral equilibrium, and the welfare optimum. For a fixed $\lambda$, we see that the consumer market size under the non-neutral equilibrium is larger than in the neutral regime. Also, it is similar to the size under the welfare optimum. Further note from Figures~\ref{fig:CP_v_eq_r} and~\ref{fig:CP_v_gr_r} that the provider market size $n$ under the neutral regime is significantly larger than under the welfare optimum and non-neutral regimes. Since both the consumer and the provider market size in the non-neutral equilibrium are closer to the welfare optimum, the non-neutral equilibrium dominates the neutral equilibrium in terms of welfare, which can be seen in Figures~\ref{fig:Wel_v_eq_r} and~\ref{fig:Wel_v_gr_r}.

Lastly, note that, in line with expectation, as $\lambda$ increases, the resulting increase in congestion at the ISP causes the size of the consumer market to steadily decrease under all three regimes. A similar trend with respect to $\lambda$ is seen for the provider market size.

We summarize the takeaway from computational analysis in the following proposition.
\begin{proposition}  \thlabel{WelfareCompare}
	[Welfare Comparison of Equilibria]. For a large selection of parameter values, the non-neutral regime is welfare superior to neutral regime.
\end{proposition}

The intuitive reason why the result holds is because a switch from a non-neutral regime to a neutral regime leads to welfare losses on both counts -- a decrease in the consumer market size that it entails decreases the economic value generated by consumers (net of any gains achieved from slightly reduced congestion costs); and an explosive increase in the provider market size (way above the welfare optimum) imposes costs on society that far outweigh any benefits to consumers from increased content diversity (number of providers).

\begin{table*}
	\small
	\begin{align}
	e(\hat{x}(d, 0), d) &= \frac{\lambda d \Big[ \mu - \lambda \big(v-d - \frac{tf}{2r} \big) + \sqrt{\big[\mu - \lambda \big(v-d - \frac{tf}{2r} \big) \big]^2 + 4 \lambda}  \Big]}{\Big[\sqrt{\big[\mu - \lambda \big(v-d - \frac{tf}{2r} \big) \big]^2 + 4 \lambda}  \Big] \Big[ \mu + \lambda \big(v-d - \frac{tf}{2r} \big) - \sqrt{\big[\mu - \lambda \big(v-d - \frac{tf}{2r} \big) \big]^2 + 4 \lambda}  \Big]}.   \label{eq3} 
	\end{align}
\end{table*}
\section{Conclusions and Future Directions}
\label{sec:conclusion}
This paper melds a competition model from the industrial organization literature with a network congestion model, in which equilibrium analysis of non-neutral and neutral regimes as well as a standard economic welfare analysis of equilibrium is carried out. We restrict ourselves to the case of a monopolistic ISP and a competitive provider market. Our results indicate a surprisingly clean welfare analysis. While there are both costs and benefits to increasing consumer and provider market size, the neutral regime is welfare dominated by the non-neutral regime. It remains to be seen whether this conclusion changes when the ISP's investment incentives are taken into account.

We focused on the market for a particular content type  with symmetric providers with the same fixed cost. However, the provider landscape is very rich, populated both by independent monopolists (big players like Google, Amazon, Facebook etc.) streaming different content types and competitors in each content type (for instance Amazon, Flipkart, Alibaba etc. in the e-commerce landscape). It is not immediately clear what is a good way to model the richness of the provider landscape. Another important extension of our work would be to the case of duopoly ISPs as is done, for instance, in \cite{economides2012network} and \cite{armstrong2006competition}. 

One difficulty with modeling is that while the neutral regime is relatively clearly defined, the non-neutral regime throws up several possibilities for the ISP. Apart from the feasibility of pricing access to the CPs, the ISP can also introduce lane prioritization and differentially price the lanes. Analyzing all the richness of possibilities in a single model seems elusive and analysis of a battery of models may be our best bet.
\appendix
\section{Appendix}
Let us define
\begin{align*}
A &= \mu + \lambda \big(v-d - \frac{t (a+f)}{2r} \big) \\
B &= \sqrt{\big[\mu - \lambda \big(v-d - \frac{t (a+f)}{2r} \big) \big]^2 + 4 \lambda}   
\end{align*}
Then, we can compute 
\begin{align*}
e(\hat{x}(d, a), d) &= \frac{\lambda d \Big[ \mu - \lambda \big(v-d - \frac{t (a+f)}{2r} \big) + B \Big]}{B(A-B)} \\
e(\hat{x}(d, a), a) &= \frac{\lambda ta \Big[ \mu - \lambda \big(v-d - \frac{t (a+f)}{2r} \big) + B \Big]}{2r B(A-B)}   
\end{align*}
Equations (\ref{FOC_d}) and (\ref{FOC_a}) can now be reduced to the following pair of equations in $(d, a)$.
\begin{align}
e(\hat{x}(d, a), d) &= \frac{da+df}{da+df+ra}   \label{eq1} \\
e(\hat{x}(d, a), a) &= \frac{f}{a+f}   \label{eq2}
\end{align}
For the neutral regime, we have~(\ref{eq3}).
Equation (\ref{eq3}) can now be used in Equation (\ref{NeutralOpt}) to solve for $d$.
\begin{spacing}{0.85}
\bibliographystyle{IEEEtran}
\bibliography{references}
\end{spacing}
\end{document}